\documentclass[sigconf,natbib=true]{acmart}

\AtBeginDocument{%
  }

\copyrightyear{2026}
\acmYear{2026}
\setcopyright{cc}
\setcctype{by-nc-nd}

\acmConference[SIGIR '26]{Proceedings of the 49th International ACM SIGIR Conference on Research and Development in Information Retrieval}{July 20--24, 2026}{Melbourne, VIC, Australia.}
\acmBooktitle{Proceedings of the 49th International ACM SIGIR Conference on Research and Development in Information Retrieval (SIGIR '26), July 20--24, 2026, Melbourne, VIC, Australia}
\acmISBN{979-8-4007-2599-9/2026/07}
\acmDOI{10.1145/3805712.3808520}

\usepackage{booktabs}
\usepackage{graphicx}
\usepackage{amsmath}
\usepackage{arydshln}   % for \hdashline in tables
\usepackage{placeins}
\usepackage{xcolor}
\usepackage{enumitem}
\usepackage{balance}

\begin{document}

\title{BEATS: Bootstrapping E-commerce Attribute Taxonomies for Search through Iterative Human-AI Collaboration}
% \titlenote{Presentation format preference: Paper, Poster, or Talk.}

%%
%% Authors and affiliations
%%
\author{Yung-Yu Shih}
\authornote{This work was done when the first author interned with Taiwan Rakuten Ichiba, Inc.}
\affiliation{%
  \institution{National Taiwan University}
  \city{Taipei}
  \country{Taiwan}
}
\email{f12944007@ntu.edu.tw}

\author{Shang-Yu Su}
\affiliation{%
  \institution{Rakuten Group, Inc.}
  \city{Tokyo}
  \country{Japan}
}
\email{shangyu.su@rakuten.com}

\author{Tzu-I Ho}
\affiliation{%
  \institution{Taiwan Rakuten Ichiba, Inc.}
  \city{Taipei}
  \country{Taiwan}
}
\email{tzui.ho@rakuten.com}

\author{Dongzhe Wang}
\affiliation{%
  \institution{Rakuten Asia Pte. Ltd.}
  \city{Singapore}
  \country{Singapore}
}
\email{dongzhe.wang@rakuten.com}

\author{Yun-Nung Chen}
\affiliation{%
  \institution{National Taiwan University}
  \city{Taipei}
  \country{Taiwan}
}
\email{y.v.chen@ieee.org}

%%
%% Short author list for page headers
\renewcommand{\shortauthors}{Yung-Yu Shih, Shang-Yu Su, Tzu-I Ho, Dongzhe Wang, and Yun-Nung Chen}

%%
%% Abstract
%%
\begin{abstract}
E-commerce platforms in emerging markets often operate with underdeveloped product catalogs that contain only category taxonomies but lack structured attribute schemas. This absence of fine-grained product attributes limits search capabilities---preventing faceted filtering, degrading query understanding, and weakening semantic representations used by search systems. We present \textbf{BEATS}, a \textbf{human-in-the-loop LLM framework} for bootstrapping product attribute taxonomies entirely from scratch. Our approach extends a multi-stage LLM generation pipeline with two critical production stages: (1) \textit{proactive quality checking} by model developers to filter erroneous outputs, and (2) \textit{human annotation} by domain-expert local staff to validate generated attributes. The framework operates iteratively---prompts at each generation stage are refined based on quality check observations and annotator feedback across successive rounds, progressively improving attribute quality. Once the attribute taxonomy is established, we employ LLMs to perform structured \textit{attribute tagging} on individual product items, enriching their contextual representations. The enriched catalog directly benefits multiple components of the search system: enabling granular attribute-based filtering, providing structured features for ranking models, and improving semantic representations for dense retrieval. We validate the generated taxonomy by training dense retrieval models on attribute-enriched product data, demonstrating consistent improvements over baselines using original catalog information. Our system has been deployed at Rakuten Taiwan, enriching 9 major categories spanning 2,694 sub-categories with 67,277 generated attributes, and over 5.4 million products have been tagged with the generated attributes, with plans to enrich the entire product catalog.
\end{abstract}

%%
%% CCS Concepts
%%
\begin{CCSXML}
<ccs2012>
  <concept>
      <concept_id>10002951.10003317.10003338</concept_id>
      <concept_desc>Information systems~Retrieval models and ranking</concept_desc>
      <concept_significance>500</concept_significance>
      </concept>
  <concept>
      <concept_id>10002951.10003317.10003347.10003352</concept_id>
      <concept_desc>Information systems~Information extraction</concept_desc>
      <concept_significance>300</concept_significance>
      </concept>
  <concept>
      <concept_id>10002951.10003260.10003282.10003550.10003555</concept_id>
      <concept_desc>Information systems~Online shopping</concept_desc>
      <concept_significance>300</concept_significance>
      </concept>
  <concept>
      <concept_id>10010147.10010178.10010187.10010195</concept_id>
      <concept_desc>Computing methodologies~Ontology engineering</concept_desc>
      <concept_significance>300</concept_significance>
      </concept>
</ccs2012>
\end{CCSXML}

\ccsdesc[500]{Information systems~Retrieval models and ranking}
\ccsdesc[300]{Information systems~Information extraction}
\ccsdesc[300]{Information systems~Online shopping}
\ccsdesc[300]{Computing methodologies~Ontology engineering}
%%
%% Keywords
%%
\keywords{Attribute Taxonomy, E-Commerce Search, Human-in-the-Loop LLM Framework, Catalog Enrichment, Dense Retrieval, Product Search}

%%
%% Build title
\maketitle

%% ============================================================================
%% SECTION 1: INTRODUCTION
%% ============================================================================
\section{Introduction}

Product catalogs are the backbone of e-commerce search systems. Well-structured catalogs with rich attribute taxonomies---such as \textit{brand}, \textit{material}, \textit{size}, and \textit{color} per product category---enable faceted navigation, precise filtering, and high-quality semantic representations~\cite{mao2020octet,zheng2018opentag}. However, e-commerce platforms in emerging markets frequently lack attribute schemas entirely, even when a basic category hierarchy exists. Building attribute taxonomies manually is prohibitively expensive across thousands of categories.

LLMs offer a promising path for automated taxonomy construction~\cite{mishra2024flame,moskvoretskii2024taxollama}, but deploying LLM-generated structured data raises quality concerns: individual LLMs exhibit biases~\cite{bender2021dangers}, and single-pass generation produces inconsistent or hallucinated attributes.

We propose \textbf{BEATS} (Bootstrapping E-commerce Attribute Taxonomies for Search), a \textbf{human-in-the-loop LLM framework} that addresses these challenges through a five-stage iterative pipeline:

\begin{enumerate}[leftmargin=*,nosep]
    \item \textbf{Multi-Source Generation:} Multiple heterogeneous LLMs independently generate candidate attributes, ensuring broad coverage through model diversity.
    \item \textbf{Generative Synthesis:} An LLM aggregates diverse candidates into a unified attribute set, resolving conflicts and merging complementary information.
    \item \textbf{Targeted Refinement:} A refinement pass removes noise and ensures consistency with the category hierarchy.
    \item \textbf{Proactive Quality Check:} Model developers review outputs to identify systematic errors before human annotation.
    \item \textbf{Human Annotation:} Domain-expert local staff validate and correct generated attributes.
\end{enumerate}

Critically, the framework operates \textit{iteratively}: insights from stages 4--5 refine prompts in stages 1--3 for subsequent rounds, progressively aligning outputs with domain-specific requirements that cannot be captured by prompting alone.

Once the taxonomy is established, we perform \textit{structured attribute tagging}: an LLM assigns attribute values to each product, enriching representations that benefit faceted filtering, ranking features, and dense retrieval. We validate effectiveness through dense retrieval experiments, demonstrating consistent improvements over baselines.

Our contributions are: (1) \textbf{BEATS}, a production-deployed \textbf{human-in-the-loop LLM framework} for bootstrapping attribute taxonomies from scratch; (2) an iterative prompt refinement methodology driven by human feedback; and (3) empirical validation that generated attributes improve search system performance.

%% ============================================================================
%% SECTION 2: RELATED WORK
%% ============================================================================
\section{Related Work}

\textbf{Taxonomy Construction and Attribute Extraction.}
Product taxonomy construction has been studied in categorization~\cite{mao2020octet} and attribute extraction~\cite{zheng2018opentag,xu2019scaling,dong2020autoknow}. Recent work leverages LLMs for taxonomy enrichment~\cite{mishra2024flame,moskvoretskii2024taxollama}, but typically assumes an existing attribute schema. Our work addresses the cold-start problem where no attribute taxonomy exists.

\textbf{LLM Ensemble and Data Generation.}
LLMs have been used for synthetic data generation~\cite{li2023syntheticdata,tang2023doessynth}, query expansion~\cite{wang2023query2doc}, and document enrichment~\cite{nogueira2019doc2query}. Multi-model ensemble approaches improve robustness~\cite{jiang2023llmblender,zhi2025llmensemble}. Our pipeline extends multi-stage LLM orchestration with human oversight for production-quality structured data.

\textbf{Human-in-the-Loop ML.}
HITL systems integrate human judgment into ML pipelines~\cite{monarch2021hitl,wu2022survey}. Active learning~\cite{settles2012active} and interactive annotation~\cite{lee2022annotation} leverage human feedback to improve outputs. Our framework differs in that humans validate \textit{structured schema} rather than individual labels, and feedback drives prompt engineering rather than model fine-tuning.

\textbf{Dense Retrieval for E-Commerce.}
Dense retrieval encodes queries and products for semantic matching~\cite{karpukhin2020dense,xiong2021ance}. Multi-aspect pre-training leverages product attributes for improved dense retrieval~\cite{sun2023multiaspect}, and generative approaches can produce structured attribute values at scale~\cite{nikolakopoulos2023sage}, but prior work assumes attributes already exist. We demonstrate that LLM-generated, human-validated attributes improve search performance.

%% ============================================================================
%% SECTION 3: METHODOLOGY
%% ============================================================================
\section{Methodology}

\subsection{Framework Overview}

Figure~\ref{fig:framework} illustrates the BEATS framework. Given a product category $c$ from the existing category taxonomy, our goal is to generate a set of validated \textit{attribute types} $\mathcal{A}_c = \{a_1, a_2, \ldots, a_k\}$ (e.g., ``material'', ``size'', ``color''). Notably, we do \textit{not} generate attribute values in the multi-stage generation pipeline; instead, the value set for each attribute is populated downstream through attribute tagging over all products under the category (Section~\ref{sec:tagging}).

The framework operates in rounds $r = 1, 2, \ldots, R$, where each round processes a different set of categories. Rather than re-generating attributes for the same categories, we apply the framework to new category sets in each round, refining prompts based on observations from prior rounds. The feedback from round $r$ informs prompt design for round $r{+}1$:
\begin{equation}
\mathcal{P}^{(r+1)} = \text{RefinePrompts}(\mathcal{P}^{(r)}, \text{QC}^{(r)}, \text{HA}^{(r)})
\end{equation}
where $\text{QC}^{(r)}$ denotes quality check observations and $\text{HA}^{(r)}$ denotes human annotation feedback from round $r$.

\begin{figure}[t]
\centering
\setlength{\tabcolsep}{2pt}
\renewcommand{\arraystretch}{1.2}
\small
\begin{tabular}{|p{7.2cm}|}
\hline
\textbf{BEATS: Human-in-the-Loop LLM Framework} \\
\hline
\textit{Round $r$:} \\[3pt]
\textbf{Stage 1: Multi-Source Generation} \\
$\{M_1, \ldots, M_n\}$ independently generate attribute candidates \\
$\mathcal{A}_c^{(i)} = M_i(\mathcal{P}_{\text{gen}}^{(r)}, c) \quad \forall i \in [n]$ \\[3pt]
\textbf{Stage 2: Generative Synthesis} \\
$M_e$ synthesizes: $\mathcal{A}_c^{(\text{syn})} = M_e(\mathcal{P}_{\text{syn}}^{(r)}, \{\mathcal{A}_c^{(1)}, \ldots, \mathcal{A}_c^{(n)}\})$ \\[3pt]
\textbf{Stage 3: Targeted Refinement} \\
$M_r$ refines: $\mathcal{A}_c^{(\text{ref})} = M_r(\mathcal{P}_{\text{ref}}^{(r)}, \mathcal{A}_c^{(\text{syn})}, c)$ \\[3pt]
\hdashline \\[-6pt]
\textbf{Stage 4: Proactive Quality Check} \\
Developers review $\mathcal{A}_c^{(\text{ref})}$, identify systematic issues \\
$\rightarrow$ Observations $\text{QC}^{(r)}$ \\[3pt]
\textbf{Stage 5: Human Annotation} \\
Domain experts validate attributes: Accept / Reject / Modify \\
$\rightarrow$ Feedback $\text{HA}^{(r)}$, Validated set $\hat{\mathcal{A}}_c$ \\[3pt]
\hdashline \\[-6pt]
\textit{Prompt Refinement:} $\mathcal{P}^{(r+1)} \leftarrow f(\mathcal{P}^{(r)}, \text{QC}^{(r)}, \text{HA}^{(r)})$ \\
\hline
\end{tabular}
\caption{The BEATS human-in-the-loop LLM framework. Stages 1--3 perform multi-stage LLM generation; Stages 4--5 provide human quality assurance. Feedback from each round refines prompts for the next, progressively improving attribute quality.}
\Description{Schematic of the five-stage BEATS pipeline. Stage~1 (Multi-Source Generation) has $n$ heterogeneous LLMs independently generating attribute candidates for a category. Stage~2 (Generative Synthesis) merges these candidates into a unified set. Stage~3 (Targeted Refinement) cleans the synthesized set against the category hierarchy. Stage~4 (Proactive Quality Check) has model developers review the refined output and produce systematic-error observations. Stage~5 (Human Annotation) has domain experts label each attribute as Accept, Unsure, or Reject. Feedback from Stages 4--5 is used to refine the prompts for Stages 1--3 in the next round.}
\label{fig:framework}
\end{figure}

\subsection{Multi-Stage LLM Generation (Stages 1--3)}

\textbf{Stage 1: Multi-Source Generation.}
We employ $n$ heterogeneous LLMs $\{M_1, \ldots, M_n\}$, deliberately chosen to maximize diversity in architecture, training data, and model scale. Each model receives a category $c$ along with its position in the category hierarchy (e.g., parent and sibling categories for context) and a generation prompt $\mathcal{P}_{\text{gen}}^{(r)}$ instructing it to produce relevant attribute types for products in this category. To facilitate cross-market knowledge transfer, prompts also incorporate reference attributes from a mature sibling platform (Rakuten Ichiba, Japan) as few-shot examples where applicable, though differences in language and catalog nature mean such references are not always directly transferable.

\textbf{Stage 2: Generative Synthesis.}
A synthesis model $M_e$ receives all candidate attribute sets and merges them into a unified schema. The synthesis prompt employs structured reasoning, instructing the model to: (1) identify unique and complementary attributes across candidates, (2) resolve naming conflicts and duplicates, (3) filter attributes that appear hallucinated or category-inappropriate, and (4) standardize attribute types.

\textbf{Stage 3: Targeted Refinement.}
A refinement model $M_r$ applies targeted de-noising. This stage cross-references the synthesized attributes against the category hierarchy to ensure consistency (e.g., attributes for ``Smartphones'' should be compatible with its parent ``Mobile Phones'') and removes attributes that are already structured at the platform level---such as price---which the catalog already exposes natively and therefore need not be regenerated as part of the attribute taxonomy.

\begin{table}[t]
\caption{Worked example for the \textit{Vacuum Cleaner} sub-category. Three representative patterns illustrate how each stage progressively addresses issues that no single LLM handles alone. ``$\rightarrow$ removed'' indicates the attribute is dropped because Stage~3 detects redundancy with another attribute or with the parent category.}
\label{tab:case-study}
\centering
\footnotesize
\renewcommand{\arraystretch}{1.2}
\setlength{\tabcolsep}{3pt}
\begin{tabular}{@{}p{1.4cm} p{2.6cm} p{1.7cm} p{1.7cm}@{}}
\toprule
Pattern & Stage~1 (3 models) & Stage~2 (synthesis) & Stage~3 (refinement) \\
\midrule
Naming variance + parent overlap &
\textit{Cordless Function} (gpt-oss); \textit{Cordless Use} (Qwen3); implicit in \textit{Power Source} (Qwen3-VL) &
\textit{Cordless Use} &
$\rightarrow$ removed (subsumed by \textit{Power Source}) \\
\hdashline
Mixed granularity &
\textit{Filter Type}; \textit{HEPA Filter}; \textit{\# Filtration Layers}; \textit{Filtration System} &
\textit{Filtration System}, \textit{Filtration Layers} &
\textit{Filtration System} only \\
\hdashline
Fragmented Yes/No tags &
\textit{Auto Suction Adjustment}; \textit{Multi-level Suction Adjustment} &
both retained as separate boolean attributes &
unified into structured \textit{Suction Control Method} \\
\bottomrule
\end{tabular}
\end{table}

\textbf{Worked Example.} We use the \textit{Vacuum Cleaner} sub-category as a running example.\footnote{Attribute names below are translated from Traditional Chinese to English for presentation; all generation, annotation, and deployment are performed natively in Traditional Chinese.}

\noindent\textit{(i)~Stage-1 model specialization.} Beyond a long shared tail of generic attributes (brand, color, weight, country of origin, \ldots), each Stage-1 LLM contributes a distinctive specialty axis that the other two miss:
\begin{itemize}[leftmargin=*,nosep]
    \item \textit{gpt-oss-120b}: unit-precise technical specs---\textit{Power}~(W), \textit{Suction Power}~(Pa), \textit{Voltage}~(V), \textit{Battery}~(mAh), \textit{Noise}~(dB)---and deeper structure (\textit{\# Filtration Layers}).
    \item \textit{Qwen3-30B}: usability and smart-control attributes (\textit{Washable Filter}, \textit{LED Light}, \textit{Auto Dust Detection}, \textit{Auto Suction Adjustment}) and use-case attributes (\textit{Pet Hair Removal}, \textit{Wet/Dry Capability}).
    \item \textit{Qwen3-VL-235B}: specific accessories (\textit{Crevice/Mattress Tool}, \textit{Wall Mount}); sanitation (\textit{UV Sterilization}, \textit{HEPA}, \textit{Mite Removal}); and an alternative suction metric \textit{Air Watts}~(AW), complementary to Pa.
\end{itemize}

\noindent\textit{(ii)~Pipeline transformations.} Table~\ref{tab:case-study} shows three representative transformations across stages. Stage~2 (synthesis) reconciles inconsistent candidates from Stage~1 (e.g., \textit{Cordless Function} vs.\ \textit{Cordless Use}; \textit{Length, Width, Height} vs.\ \textit{Dimensions}) but tends to retain near-duplicate or overly fine-grained candidates because it has no schema-level view of the catalog. Stage~3 (refinement) then drops redundancy already implied by a parent attribute (\textit{Cordless Use} $\subset$ \textit{Power Source}), collapses over-granular variants (\textit{Filtration Layers} into \textit{Filtration System}), and converts fragmented Yes/No flags into a unified categorical attribute (\textit{Suction Control Method})---the last is what enables a clean faceted-search UI rather than a long list of boolean checkboxes.

\textbf{Cross-Market Knowledge Transfer.}
A distinctive feature of our setting is the asymmetry between two sibling platforms operated by the same group: Rakuten Ichiba (Japan) has a mature attribute taxonomy curated over many years, while Rakuten Taiwan has only a category hierarchy. We exploit this asymmetry by injecting the Japan attribute set for the parent category as few-shot references in Stage~1 prompts---but as soft guidance, not ground truth. Naive transfer is ineffective for three reasons: (i) the two platforms operate in different languages with different writing systems (Japanese kanji/kana vs.\ Traditional Chinese), so attribute names cannot be transliterated verbatim; (ii) the Taiwan catalog skews toward different sub-segments than Japan (e.g., a distinct fashion mix and different consumer-electronics brand portfolios), exposing attribute gaps that the Japan reference does not cover; and (iii) some Japan-specific attributes---locale-specific units, regional certifications, market-specific delivery formats---are simply inapplicable in Taiwan. A subtler failure mode arises precisely \textit{because} Japanese and Traditional Chinese share a large set of Han characters: Stage-1 LLMs occasionally emit attributes that read fluently in Traditional Chinese yet smuggle in Japan-specific concepts. For example, our Stage-4 proactive quality check caught a generated attribute meaning ``Origin (\textit{todōfuken})''---which imposes Japan's four-character prefecture system---when the correct Taiwan form is the bare ``Origin'', since Taiwan's administrative divisions do not follow the \textit{todōfuken} scheme. Such bleed-through is invisible at the surface-fluency level and is exactly the cross-market error class that the human-in-the-loop validation in Stages~4--5 is designed to surface.

\subsection{Human Quality Assurance (Stages 4--5)}

\textbf{Stage 4: Proactive Quality Check.}
Before engaging domain-expert annotators, model developers perform systematic quality reviews. This stage identifies \textit{systematic} issues rather than individual errors---for example, the model consistently generating overly technical attributes for consumer-facing categories, or missing critical attributes for certain category types. These observations inform targeted prompt adjustments.

\textbf{Stage 5: Human Annotation.}
Domain-expert staff label each attribute as \textit{Accept} (correct and relevant), \textit{Unsure} (applicability unclear), or \textit{Reject} (incorrect or irrelevant). Each attribute is independently judged by multiple annotators with majority vote; the small fraction of attributes with no emerging majority (i.e., 1-1-1 splits) is conservatively treated as \textit{Unsure}. Results serve dual purposes: producing the validated taxonomy for deployment and providing feedback signals for prompt refinement.

\subsection{Iterative Prompt Refinement}

After each round, we analyze quality check observations and annotation statistics to refine prompts. Common refinements include adding negative examples of rejected attributes, adjusting granularity instructions, incorporating category-specific guidance, and tuning the refinement stage to catch recurring error types.

\subsection{Downstream: Attribute Tagging and Search}
\label{sec:tagging}

\textbf{Attribute Tagging.}
Once the validated attribute taxonomy $\hat{\mathcal{A}}_c$ (attribute types) is established for category $c$, an LLM tags each product item $d$ by predicting values for applicable attributes given the item's title and description, producing $d^* = d \oplus \text{Tag}(d, \hat{\mathcal{A}}_c)$. The aggregated predicted values across products naturally form each attribute's value set, eliminating the need to pre-define values during taxonomy generation.

\textbf{Search System Impact and Validation.}
The enriched catalog benefits: (1) \textit{Faceted Search} via attribute filters previously unavailable; (2) \textit{Ranking Features} for learning-to-rank; and (3) \textit{Dense Retrieval} through improved semantic matching. We validate by training bi-encoder dense retrieval models~\cite{karpukhin2020dense} on both original and attribute-enriched items, using retrieval performance as a proxy for taxonomy quality.

%% ============================================================================
%% SECTION 4: EXPERIMENTS
%% ============================================================================
\section{Experiments}

\subsection{Setup}

\textbf{Platform and Data.}
We deploy BEATS on Rakuten Taiwan, a major e-commerce platform whose catalog contains a category hierarchy but lacks attribute schemas entirely. We select 9 major categories spanning 2,694 sub-categories as our experimental scope, covering a substantial portion of the platform's product offerings. All data---category names, product titles, and descriptions---is in Traditional Chinese, presenting additional challenges for LLM generation compared to English-centric benchmarks. In total, 67,277 attributes are generated and over 5.4 million products are tagged.

\textbf{LLM Configuration.}
We conduct two rounds of generation over different category sets. For the generation stage (Stage 1), we employ three heterogeneous LLMs in both rounds: GPT-OSS-120B, Qwen3-30B-A3B-Instruct, and Qwen3-VL-235B-A22B-Instruct-FP8, spanning proprietary and open-source families with varying scales and architectures. These three were chosen after pilot studies over a broader pool of well-known open-source LLMs across different scales: we shortlisted the models that produced the strongest attribute-generation outputs on our Traditional Chinese e-commerce data, since performance on this language and task was a more reliable indicator than general-purpose benchmark scores. For synthesis (Stage 2), both rounds use Qwen3-30B-A3B-Instruct. For refinement (Stage 3), Round 1 uses Qwen3-30B-A3B-Instruct while Round 2 uses GPT-OSS-120B, as quality check observations from Round 1 indicated that a stronger refinement model reduced systematic errors. Generation prompts also incorporate category and attribute information from Rakuten Ichiba (Japan) as few-shot references for cross-market knowledge transfer, though differences in language and catalog nature limit direct transferability. All LLM inference is served via vLLM~\cite{kwon2023vllm}.

\textbf{Human Annotation Setup.}
Quality checks (Stage 4) are performed by model developers. Human annotation (Stage 5) involves 11 local domain-expert staff at Rakuten Taiwan who are familiar with the platform's product categories and consumer behavior. Each generated attribute is independently annotated by 3 annotators, who choose from: (1) \textit{Accept}, (2) \textit{Unsure}, or (3) \textit{Reject}. The final decision is determined by majority vote (at least 2 out of 3). In Round 2, we additionally ask annotators: ``Do you think the quality of the generated attributes is better in this round?'' with choices (1) Yes, (2) Same, (3) No, to directly validate the effectiveness of iterative prompt refinement.

\textbf{Dense Retrieval Setup.}
We implement a bi-encoder dense retrieval model with a BERT-based backbone~\cite{devlin2019bert} using PyTorch and HuggingFace Transformers, trained with hard negative mining~\cite{xiong2021ance}. We compare:
\begin{itemize}[leftmargin=*,nosep]
    \item \textbf{Baseline}: Original item information (title + description)
    \item \textbf{+Attributes}: Item information enriched with tagged attributes from our generated taxonomy
\end{itemize}

\textbf{Metrics.}
Taxonomy quality: acceptance rate, unsure rate, rejection rate per round; Round 2 perceived quality improvement.
Retrieval: Recall@10, Recall@100, NDCG@10, MRR@10.

\subsection{Taxonomy Generation Quality}

% TODO: Fill with actual results
\begin{table}[t]
\caption{Attribute quality across iteration rounds (majority vote over 3 annotators per attribute). Round 2 processes different categories with refined prompts.}
\label{tab:iteration}
\centering
\begin{tabular}{lccc}
\toprule
Round & Accept (\%) & Unsure (\%) & Reject (\%) \\
\midrule
Round 1 & 89.00 & 7.60 & 3.43 \\
Round 2 & \textbf{99.88} & 0.10 & 0.015 \\
\bottomrule
\end{tabular}
\end{table}

Table~\ref{tab:iteration} presents the attribute quality across iteration rounds, determined by majority vote of 3 independent annotators. The acceptance rate improves dramatically from 89.00\% in Round 1 to 99.88\% in Round 2, while the rejection rate drops from 3.43\% to near zero (0.015\%). This substantial improvement validates the effectiveness of iterative prompt refinement: quality check observations from Round 1 identified systematic error patterns (e.g., overly generic attributes, language-specific issues in Traditional Chinese generation), which informed targeted prompt adjustments and the upgrade of the refinement model from Qwen3-30B to GPT-OSS-120B in Round 2.

In Round 2, annotators are additionally asked whether the generated attributes are of higher quality compared to Round 1. All annotators (100\%) responded ``Yes'', providing strong direct human validation that the iterative prompt refinement process yields tangible quality improvements perceived by domain experts.

\begin{table}[t]
\caption{Inter-annotator agreement breakdown across the 3 annotators per attribute. ``3/3'' denotes unanimous agreement; ``2-1'' denotes a majority of 2 with one dissent; ``1-1-1'' denotes full disagreement (no majority).}
\label{tab:iaa}
\centering
\begin{tabular}{lcccc}
\toprule
Round & 3/3 (\%) & 2-1 (\%) & 1-1-1 (\%) & Fleiss' $\kappa$ \\
\midrule
Round 1 & 73.27 & 23.13 & 3.60 & 0.24 \\
Round 2 & \textbf{96.34} & 3.57 & \textbf{0.09} & 0.02$^{\dagger}$ \\
\bottomrule
\end{tabular}
\end{table}

Table~\ref{tab:iaa} reports the inter-annotator agreement breakdown. Round 2 exhibits substantially higher unanimous (3/3) agreement (96.34\% vs.\ 73.27\%) and an order-of-magnitude lower full-disagreement (1-1-1) rate (0.09\% vs.\ 3.60\%), directly evidencing the perceived quality improvement and the reduced ambiguity of Round-2 attributes. We note that the lower Round-2 Fleiss' $\kappa$ ($^{\dagger}$) reflects the well-known paradox of $\kappa$ under extreme class imbalance: when one category overwhelmingly dominates (here \textit{Accept}~$>$99\%), the expected-by-chance agreement approaches the observed agreement, so $\kappa$ collapses despite the actual rater agreement being substantially higher.

\subsection{Dense Retrieval Results}

\begin{table}[t]
\caption{Dense retrieval results comparing original vs.\ attribute-enriched item representations on Rakuten Taiwan product search.}
\label{tab:retrieval}
\centering
\begin{tabular}{lcccc}
\toprule
Setting & R@10 & R@100 & NDCG@10 & MRR@10 \\
\midrule
Baseline & 66.6 & 83.8 & 49.2 & 44.5 \\
+Attributes & \textbf{70.4} & \textbf{86.8} & \textbf{52.0} & \textbf{46.9} \\
\bottomrule
\end{tabular}
\end{table}

Table~\ref{tab:retrieval} shows dense retrieval results. Attribute-enriched item representations consistently outperform the baseline across all metrics. Recall@10 improves from 66.6 to 70.4 (+5.7\%) and Recall@100 from 83.8 to 86.8 (+3.6\%), demonstrating that the generated attribute taxonomy captures meaningful product semantics that help the dense retrieval model better match user queries. NDCG@10 improves from 49.2 to 52.0 (+5.7\%) and MRR@10 from 44.5 to 46.9 (+5.4\%), indicating substantially more accurate top-ranked results with attribute enrichment. These improvements are achieved purely through enriching item text representations with LLM-generated and human-validated attributes, without any changes to the retrieval model architecture or training procedure.

\subsection{Attribute Tagging Quality}

Since the attribute tagging step operates on the generated taxonomy, the quality of tagging results indirectly reflects the quality of the underlying attribute taxonomy. We evaluate tagging quality using an LLM-as-a-judge approach~\cite{zheng2023judging}: for each predicted attribute type-value pair assigned to a product, SOLAR-10.7B-Instruct~\cite{kim2024solar} (served via vLLM) assesses whether the pair is appropriate for the given product, yielding a quality score. Importantly, the judge model is drawn from a different model family than any LLM used in generation, synthesis, refinement, or tagging (Qwen3 / GPT-OSS), so it provides an independent assessment rather than self-evaluation; SOLAR was chosen because prior work reports its judgments correlate strongly with GPT-4 in LLM-as-a-judge settings~\cite{rau2024bergen}, while remaining cheap enough to scale on our infrastructure. Due to the computational cost of LLM-based evaluation, we evaluate on stratified samples: 9,717 tagged products from Round 1 categories and 20,900 from Round 2 categories.

\begin{table}[t]
\caption{Attribute tagging quality evaluated by LLM-as-a-judge on sampled products. Scores reflect the appropriateness of predicted attribute type-value pairs, indirectly validating the generated taxonomy quality.}
\label{tab:tagging}
\centering
\begin{tabular}{lcc}
\toprule
Category Scope & \#Samples & Score \\
\midrule
Round 1 categories (3 major) & 9,717 & 0.926 \\
Round 2 categories (6 major) & 20,900 & \textbf{0.936} \\
\hdashline
Overall (all 9 major) & 30,617 & 0.933 \\
\bottomrule
\end{tabular}
\end{table}

Table~\ref{tab:tagging} presents the tagging quality scores. Round 2 categories achieve a higher score (0.936) than Round 1 categories (0.926), consistent with the improved taxonomy quality from iterative prompt refinement. The overall score of 0.933 across all 9 major categories indicates that the vast majority of generated attribute type-value pairs are judged appropriate by the LLM judge, further validating both the taxonomy generation and the attribute tagging pipeline.

%% ============================================================================
%% SECTION 5: DEPLOYMENT AND LESSONS LEARNED
%% ============================================================================
\section{Deployment and Lessons Learned}

\textbf{Production Impact.}
BEATS has been deployed at Rakuten Taiwan, enriching 9 major categories spanning 2,694 sub-categories with 67,277 generated attributes. Over 5.4 million products have been tagged with structured attribute information, with plans to extend coverage to the entire product catalog.

\textbf{Practical Lessons.}
\begin{itemize}[leftmargin=*,nosep]
    \item \textit{Proactive quality checks are essential.} Without Stage 4, annotators were overwhelmed by systematic errors that could have been caught earlier, reducing annotation efficiency.
    \item \textit{Cross-market knowledge transfer has limits.} Reference attributes from Rakuten Ichiba (Japan) provided useful few-shot context, but language differences (Japanese vs.\ Traditional Chinese) and catalog nature differences required careful prompt adaptation.
    \item \textit{Category hierarchy context matters.} Providing parent and sibling category information in prompts significantly reduced out-of-scope attribute generation.
    \item \textit{Stronger refinement models help.} Upgrading the refinement model in Round 2 noticeably reduced residual noise, suggesting refinement benefits more from model capability than generation does.
    \item \textit{Taxonomy quality drives tagging quality.} Downstream tagging errors were primarily traceable to taxonomy issues, validating our focus on generation quality.
\end{itemize}

%% ============================================================================
%% SECTION 6: CONCLUSION
%% ============================================================================
\section{Conclusion}

We presented BEATS, a human-in-the-loop LLM framework that bootstraps product attribute taxonomies from scratch by combining multi-stage LLM generation with proactive quality checking and domain-expert annotation, with prompts iteratively refined from human feedback. Deployed at Rakuten Taiwan with 67K attributes across 9 major categories and 5.4M+ tagged products, BEATS yields production-quality schemas that enable faceted filtering, supply structured features for ranking, and improve dense-retrieval representations. Future work includes active-learning strategies for more efficient annotation, automatic attribute-value hierarchy generation, and applying enrichment to query understanding and learning-to-rank.

%% ============================================================================
%% INDUSTRY TRACK REQUIRED: PRESENTER BIO
%% (does not count towards page limit per submission guidelines)
%% ============================================================================
\section*{Presenter Bio}
\noindent Shang-Yu Su is a principal research scientist at Rakuten, where he works on search, including retrieval and ranking models. His research interests also span dialogue systems and natural language processing. He has 10 years of experience in NLP and deep learning and has published at venues including ACL and EMNLP. He was a Google Fellowship recipient. He holds a PhD from National Taiwan University.

%%
%% Acknowledgments - uncomment for camera-ready
%%
% \begin{acks}
% This work was supported by ...
% \end{acks}

%% Force all floating tables/figures to be placed before references
\FloatBarrier

%%
%% Bibliography
%%
\bibliographystyle{ACM-Reference-Format}
\balance
\bibliography{references}

\end{document}